\documentclass[letter]{emulateapj-rtx4}
\usepackage[usenames,dvipsnames]{color}
\usepackage{natbib,graphicx,latexsym,lscape,pdflscape,url,pdfpages}
\usepackage[caption=false]{subfig}
\usepackage{lscape}
\newcommand{\Hipparcos}{{\sl Hipparcos}}
\newcommand{\Gaia}{{\sl Gaia}}

\newcommand{\Msun}{\mbox{$M_{\sun}$}}

\newcommand{\Lsun}{\mbox{$L_{\sun}$}}

\newcommand{\Mjup}{\mbox{$M_{\rm Jup}$}}

\newcommand{\degree}{\mbox{$^{\circ}$}}

\newcommand{\kms}{\mbox{km\,s$^{-1}$}}
\newcommand{\masyr}{\hbox{mas\,yr$^{-1}$}}

\newcommand{\Lbol}{\mbox{$L_{\rm bol}$}}

\newcommand{\bpic}{$\beta$~Pic}

\begin{document}

\title{A Model-Independent Mass and Moderate Eccentricity for $\beta$~Pic~b}

\author{Trent J.\ Dupuy,\altaffilmark{1} 
        Timothy D.~Brandt,\altaffilmark{2},
        Kaitlin Kratter\altaffilmark{3},
        and
        Brendan P.~Bowler\altaffilmark{4}}

      \altaffiltext{1}{Gemini Observatory, Northern Operations Center,
        670 N.\ A'ohoku Place, Hilo, HI 96720, USA}

      \altaffiltext{2}{Department of Physics, University of California, Santa Barbara, Santa Barbara, CA 93106, USA}

      \altaffiltext{3}{Department of Astronomy and Steward Observatory, University of Arizona, Tucson, AZ 85721, USA}

      \altaffiltext{4}{The University of Texas at Austin, Department
        of Astronomy, 2515 Speedway C1400, Austin, TX 78712, USA}

\begin{abstract}

We use a cross-calibration of \Hipparcos\ and \Gaia~DR2 astrometry for \bpic\ to measure the mass of the giant planet \bpic~b ($13\pm3$\,\Mjup) in a comprehensive joint orbit analysis that includes published relative astrometry and radial velocities. Our mass uncertainty is somewhat higher than previous work because our astrometry from the \Hipparcos--\Gaia\ Catalog of Accelerations accounts for the error inflation and systematic terms that are required to bring the two data~sets onto a common astrometric reference frame, and because we fit freely for the host-star mass ($1.84\pm0.05$\,\Msun). This first model-independent mass for a directly imaged planet is inconsistent with cold-start models given the age of the \bpic\ moving group ($22\pm6$\,Myr) but consistent with hot- and warm-start models, concordant with past work. We find a higher eccentricity ($0.24\pm0.06$) for \bpic~b compared to previous orbital fits. If confirmed by future observations, this eccentricity may help explain inner edge, scale height, and brightness asymmetry of \bpic's disk. It could also potentially signal that \bpic~b has migrated inward to its current location, acquiring its eccentricity from interaction with the 3:1 outer Lindblad resonance in the disk.

\end{abstract}

\keywords{astrometry --- planetary systems --- stars: individual (bet Pic)}


\section{Introduction}

Directly imaged planets typically have their masses inferred indirectly from their luminosity and age, using uncalibrated evolutionary models that assume an initial thermal state. Most commonly-used models assume an initially high specific entropy \citep[hot start; e.g.,][]{1997ApJ...491..856B,2003A&A...402..701B}, but the planet formation process might radiate away a significant amount of energy leading to a much lower initial specific entropy \citep[cold or warm start; e.g.,][]{2007ApJ...655..541M,2012ApJ...745..174S}. Furthermore, planet assembly could be slow and only conclude well after the star is formed, in which case young planets could appear even more luminous than hot-start models would predict from the host star's age. The crucial observations needed to sort out these various possibilities are masses of planets with known age and luminosity.

\bpic~b was one of the first directly imaged planets to be discovered \citep{2010Sci...329...57L}, and its host star is the namesake of a young moving group of well-determined age \citep[$22\pm6$\,Myr; e.g.,][]{2014MNRAS.438L..11B,2017AJ....154...69S}. We present here a new model-independent dynamical mass for \bpic~b. We use the methodology of \citet{Brandt_Dupuy_Bowler_2018} to perform a joint orbital analysis of relative astrometry, radial velocities, and host-star astrometry from the cross-calibrated \Hipparcos--\Gaia\ Catalog of Accelerations \citep[HGCA;][]{Brandt_2018}.  Our new mass is consistent with recent results from \citet{2018NatAs.tmp..114S} but with broader uncertainties owing to our re-assessment of errors reported in \Hipparcos\ and \Gaia~DR2 catalogs.

\section{Data} \label{sec:data}

\subsection{Host-Star Astrometry}

\citet{Brandt_2018} has cross-calibrated \Hipparcos\ and \Gaia~DR2, placing them on a common reference frame.
Figure~1 of \cite{Brandt_2018} shows that neither the \Hipparcos\ re-reduction \citep{2007A&A...474..653V} nor the \Gaia~DR2 astrometry \citep{2018A&A...616A...2L} are suitable for orbit fitting in their published form: the ensemble of proper motion differences are inconsistent with their formal uncertainties.
Moreover, Figure~9 of \cite{Brandt_2018} shows that the cross-calibrated HGCA proper motions satisfy the standard assumptions of Gaussianity but that the lowest-precision stars in \Gaia\ (like \bpic) have uncertainties that remain underestimated.

HGCA contains three proper motions: a (nearly) instantaneous proper motion near 1991.25, another near 2015.5, and the positional difference between the catalogs scaled by the time between them.  The three proper motions are nearly independent.
\citet{Brandt_2018} also gives the central epoch at which a position was measured; this is the epoch with the minimum positional uncertainty (which differs slightly in right ascension and declination).

\begin{deluxetable*}{lccccccr}
\tablewidth{0pt}
\tablecaption{Absolute Stellar Astrometry}
\tablehead{
    \colhead{Mission} &
    \colhead{$\mu_{\alpha*}$} &
    \colhead{$\sigma[\mu_{\alpha*}]$} &
    \colhead{$\mu_{\delta}$} &
    \colhead{$\sigma[\mu_\delta]$} &
    \colhead{Corr$[\mu_{\alpha*},\mu_\delta]$} &
    \colhead{$t_{\alpha*}$} &
    \colhead{$t_\delta$} \\
    \colhead{} &
    \multicolumn{2}{c}{(mas\,yr$^{-1}$)} & 
    \multicolumn{2}{c}{(mas\,yr$^{-1}$)} &
    \colhead{} &
    \multicolumn{2}{c}{(year)}
    }
\startdata
\Hipparcos        & 4.4   & 0.4                  & 82.8   & 0.4                  & 0.002 & 1991.33 & 1991.26 \\
\Hipparcos--\Gaia & 4.796 & 0.027                & 83.863 & 0.028                & 0.025 & \nodata & \nodata \\
\Gaia             & 2.5   & 2.5\tablenotemark{a} & 82.6   & 2.5\tablenotemark{a} & 0.040 & 2015.58 & 2015.67
\enddata
\tablenotetext{a}{\Gaia~DR2 errors have been inflated by a factor of two as recommended by \cite{Brandt_2018} for stars like \bpic\ that have large reported proper motion errors in DR2 ($\gtrsim0.7$\,\masyr).}
\label{tab:hip_gaia}
\end{deluxetable*}

Table \ref{tab:hip_gaia} lists our HGCA proper motions for \bpic, the correlation coefficients between proper motion in RA and Dec, and the central epoch for each measurement. \bpic\ is heavily saturated in \Gaia\ data and thus is among the least-precisely measured stars in the HGCA.  Figure~9 of \citet{Brandt_2018} indicates that the inflated uncertainties of such stars remain underestimated by as much as factor of two.  We have therefore doubled the \Gaia~DR2 proper motion errors beyond the values in the HGCA.  For parallax, we adopt the same 60/40 linear combination of the \Hipparcos\ catalogs as the HGCA and add the same 0.20\,mas error inflation in quadrature; this results in a value of $51.61\pm0.39$\,mas.

The uncertainties for the \Hipparcos\ proper motions are much larger in the \cite{Brandt_2018} catalog than in the \Hipparcos\ re-reduction, though they are slightly smaller than the uncertainties of the original \Hipparcos\ reduction.  This is a generic feature of bright stars in the HGCA.  As shown in Figure~1 of \cite{Brandt_2018}, stars with higher-precision proper motions depart most strongly from the standard normal distribution in their residuals.  Even for the most precise 20\% of stars, a $\sim$60/40 linear combination of the two \Hipparcos\ reductions gives lower residuals than the \citet{2007A&A...474..653V} proper motions alone, and further error inflation is necessary to bring the residuals into agreement with a normal distribution. Given the cross-calibration approach used in the HGCA, it would be infeasible to use \Hipparcos\ epoch astrometry, as in \citet{2018NatAs.tmp..114S}, and ensure independence of individual measurements.

\subsection{Literature Relative Astrometry \& Radial Velocities}

We consider all available relative astrometry of \bpic~b in our orbit analysis, setting aside duplicate measurements when the same data have been analyzed separately in the literature. This includes astrometry from VLT/NaCo \citep{2010ApJ...722L..49Q,2011A&A...528L..15B,2013A&A...555A.107B,2012A&A...542A..41C,2013A&A...559L..12A,2014A&A...566A..91M}, Gemini-S/NICI \citep{2014ApJ...794..158N}, Magellan/MagAO \citep{2014ApJ...794..158N}, Gemini-S/GPI \citep{2016AJ....152...97W}, and VLT/SPHERE \citep{2018arXiv180908354L}. This comprises 50 measurements spanning sixteen years, with two observations on the northeastern side of the orbit (in 2003~November and 2018~September).

The radial velocity of the host star has been monitored from 2003--2011 with the HARPS spectrograph \citep{2012A&A...542A..18L}. We use all 1049 individual published measurements and account for the substantial intrinsic ``jitter'' that is expected for a young star like \bpic. We also use the measurement of the planet's relative radial velocity ($\Delta{\rm RV} = {\rm RV}_{\rm comp} - {\rm RV}_{\rm host}$) from \citet{2014Natur.509...63S} in our orbit fit.

\section{Orbit Analysis} \label{sec:orbit}

Relative astrometry from direct imaging has already been shown to constrain many orbital parameters of \bpic~b given the long time baseline and intensive monitoring \citep[e.g.,][]{2016AJ....152...97W,2018arXiv180908354L}. Therefore, as a first step we fit the relative astrometry with a standard seven-parameter Keplerian orbit in order to assess any systematics in combining astrometry from many different instruments and data reduction methods. We found an unreasonably large $\chi^2$ of 165 for 93 degrees of freedom (dof), $p(\chi^2) = 6\times10^{-6}$, when taking all reported astrometric errors at face value. To achieve $p(\chi^2) = 0.5$ we estimated that errors of 4\,mas and 0$\fdg$3 would need to be added in quadrature to all separation and PA measurements, respectively. Alternatively, we could exclude a handful of outlier measurements (which have reasons for being suspect) to decrease the $\chi^2$ of the maximum likelihood solution to a reasonable value. 

Five epochs of VLT/NaCo astrometry from \citet{2014A&A...566A..91M} account for 30\% of the $\chi^2$ in the relative orbit fit. \citet{2012A&A...542A..41C} and \citet{2018arXiv180908354L} did not use any of these five epochs, even though they each could have used at least some.  We therefore exclude all \citet{2014A&A...566A..91M} astrometry, which is contemporaneous with other available measurements. Likewise, Gemini/NICI astrometry from \citet{2014ApJ...794..158N} has three highly discrepant measurements (25\% of the total $\chi^2$) that each were obtained on the same night as another measurement that is more consistent with the orbit fit. We exclude these three measurements as well, using 42 relative astrometry measurements in our final orbital analysis.

As shown by \cite{Brandt_Dupuy_Bowler_2018}, simultaneous measurements of projected relative separation, host-star radial velocity, and host-star astrometric acceleration can provide a direct measurement of companion mass. In practice, observations of directly imaged companions are never truly simultaneous, although for very long long orbital periods on the order of centuries this can be a good approximation. The orbit of \bpic~b is of order decades \citep[e.g.,][]{2012A&A...542A..41C,2014ApJ...794..158N}, so more detailed analysis is needed to produce a companion mass from combining these three types of measurements. Our approach is described in detail in \cite{Brandt_Dupuy_Bowler_2018} and briefly here.

\begin{figure*}
\centerline{
\includegraphics[width=0.30\linewidth]{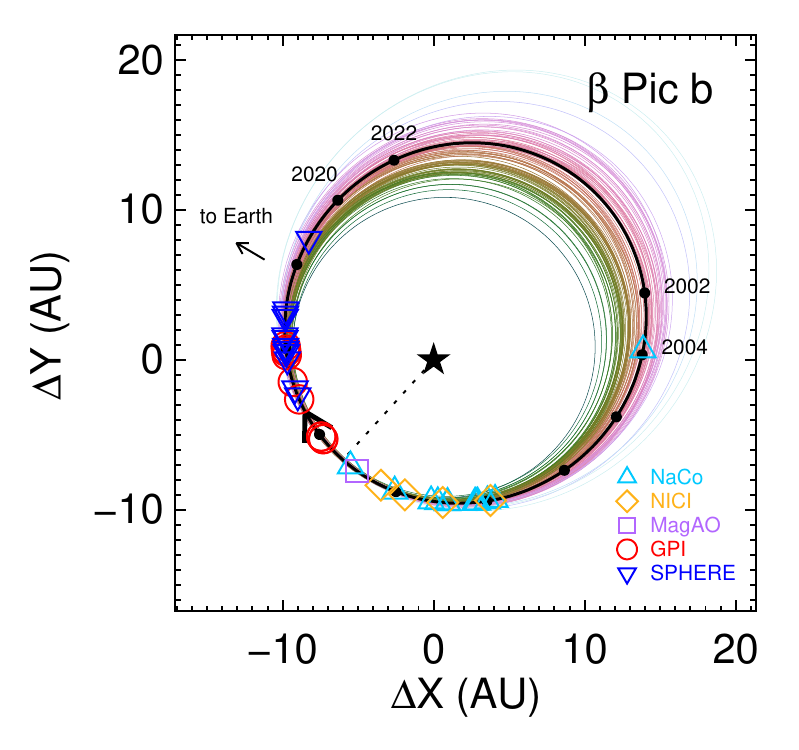}
\hskip -0.1 truein
\includegraphics[width=0.30\linewidth]{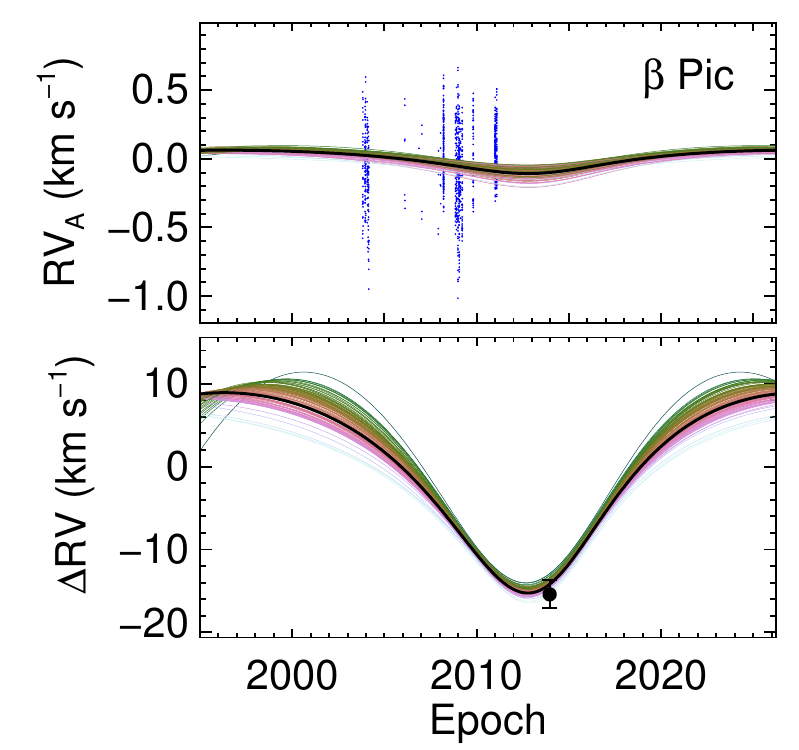}
\hskip -0.1 truein
\includegraphics[width=0.30\linewidth]{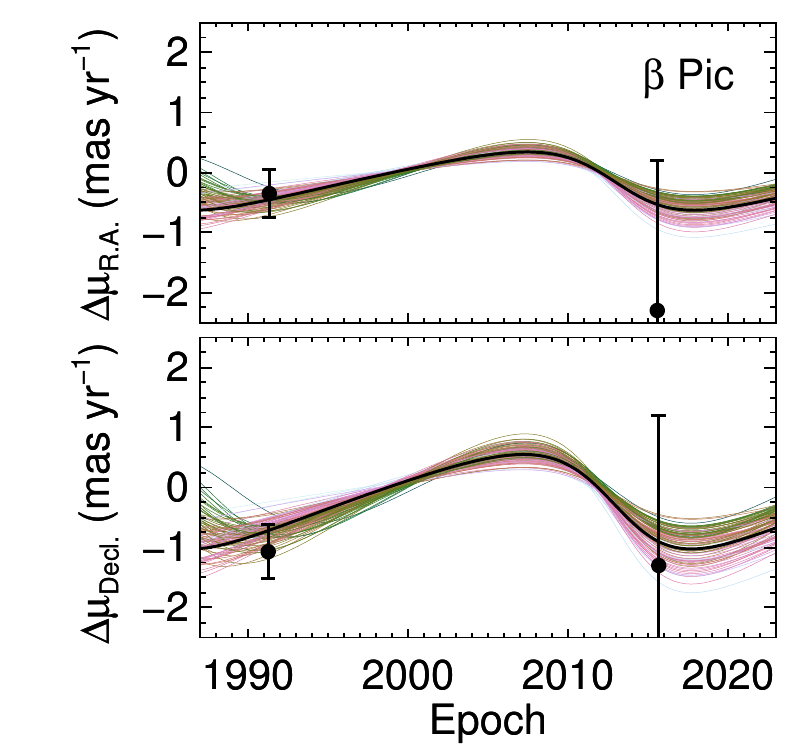}
\hskip -1.3 truein
\includegraphics[width=0.30\linewidth]{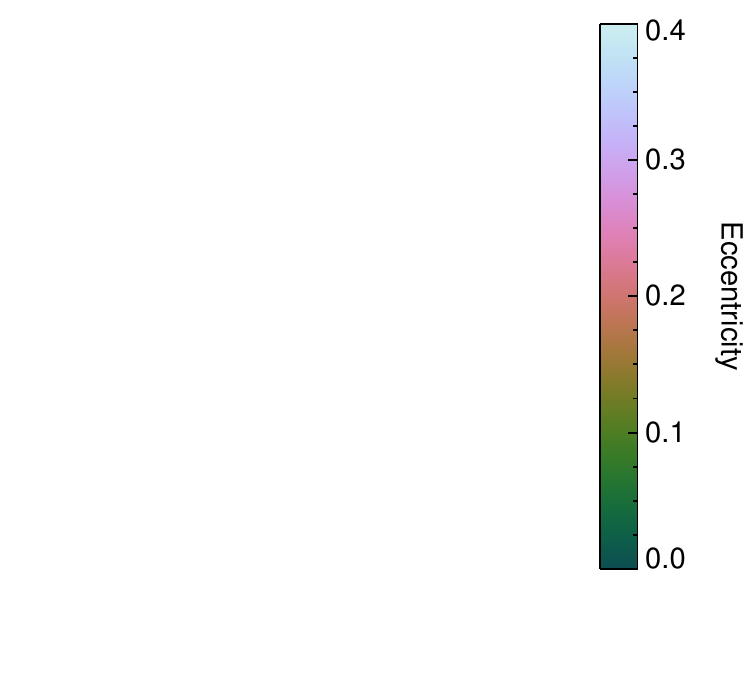}
}
\vskip 0.0 truein
\caption{Our joint orbit fit to relative astrometry (left), RVs (middle), and absolute astrometry of the host star from HGCA (right). In all panels, the thick black line indicates the highest likelihood orbit, and thin lines are 100~orbits drawn randomly from our posterior distribution colored according to orbital eccentricity.
{\bf Left:} Small filled circles along the maximum likelihood deprojected orbit indicate epochs spaced by 2~years from 2002 until 2022. The dotted line indicates periastron. Open symbols of different shapes and colors are plotted along the maximum likelihood orbit at the epochs corresponding to the relative astrometry used in our analysis.
{\bf Middle:} Over $10^3$ RVs for \bpic\ from \citet{2012A&A...542A..18L} are plotted as small blue dots, displaying a large jitter of $269\pm6$\,m\,s$^{-1}$. The bottom panel shows \bpic~b's RV relative to its host star along with the measurement of $-15.4\pm1.7$\,\kms\ from \citet{2014Natur.509...63S}. 
{\bf Right:} Each plotted measurement is the difference between the proper motion measured in one mission (\Hipparcos\ in 1991.3 or \Gaia\ in 2015.6) and the proper motion computed from the change in RA and Dec between the two missions. The strongest constraint on acceleration caused by \bpic~b comes from \Hipparcos\ given the large astrometric errors for \bpic\ in \Gaia~DR2.}
\label{fig:orbit}
\end{figure*}

\begin{deluxetable*}{lccc}
\tablecaption{MCMC Orbital Posteriors for $\beta$ Pic b \label{tbl:mcmc-BPIC}}
\setlength{\tabcolsep}{0.10in}
\tabletypesize{\tiny}
\tablewidth{0pt}
\tablehead{
\colhead{Property}              &
\colhead{Median $\pm$1$\sigma$} &
\colhead{95.4\% c.i.}           &
\colhead{Prior}                 }
\startdata
\multicolumn{4}{c}{Fitted parameters} \\[1pt]
\cline{1-4}
\multicolumn{4}{c}{} \\[-5pt]
Companion mass $M_{\rm comp}$ (\Mjup)                                       & $13.1_{-3.2}^{+2.8}$             &          7.2, 19.5         & $1/M$ (log-flat)                                                   \\[3pt]
Host-star mass $M_{\rm host}$ (\Msun)                                       & $1.84\pm0.05$                    &         1.74, 1.94         & $1/M$ (log-flat)                                                   \\[3pt]
Parallax (mas)                                                              & $51.60_{-0.39}^{+0.40}$          &        50.82, 52.37        & $\exp[-0.5((\varpi-\varpi_{\rm DR2})/\sigma[\varpi_{\rm DR2}])^2]$ \\[3pt]
Semimajor axis $a$ (AU)                                                     & $11.8_{-0.9}^{+0.8}$             &         10.3, 13.7         & $1/a$ (log-flat)                                                   \\[3pt]
Inclination $i$ (\degree)                                                   & $88.87\pm0.08$                   &        88.71, 89.04        & $\sin(i)$, $0\degree < i < 180\degree$                             \\[3pt]
$\sqrt{e}\sin{\omega}$                                                      & $-0.080_{-0.029}^{+0.027}$       &     $-$0.134, $-$0.017     & uniform                                                            \\[3pt]
$\sqrt{e}\cos{\omega}$                                                      & $-0.48\pm0.05$                   &      $-$0.59, $-$0.36      & uniform                                                            \\[3pt]
Mean longitude at $t_{\rm ref}=2455197.5$~JD, $\lambda_{\rm ref}$ (\degree) & $150\pm4$                        &          142, 159          & uniform                                                            \\[3pt]
PA of the ascending node $\Omega$ (\degree)                                 & $31.65\pm0.09$                   &        31.48, 31.82        & uniform                                                            \\[3pt]
RV zero point (m\,s$^{-1}$)                                      & $73_{-15}^{+14}$                 &           45, 103          & uniform                                                            \\[3pt]
RV jitter $\sigma$ (m\,s$^{-1}$)                                 & $269\pm6$                        &          257, 281          & $1/\sigma$ (log-flat)                                              \\[3pt]
\cline{1-4}
\multicolumn{4}{c}{} \\[-5pt]
\multicolumn{4}{c}{Computed properties} \\[1pt]
\cline{1-4}
\multicolumn{4}{c}{} \\[-5pt]
Orbital period $P$ (yr)                                                     & $29.9_{-3.2}^{+2.9}$             &         24.1, 36.8         & \nodata                                                            \\[3pt]
Semimajor axis (mas)                                                        & $610_{-50}^{+40}$                &          530, 700          & \nodata                                                            \\[3pt]
Eccentricity $e$                                                            & $0.24\pm0.06$                    &         0.13, 0.35         & \nodata                                                            \\[3pt]
Argument of periastron $\omega$ (\degree)                                   & $189.3_{-2.9}^{+3.0}$            &        182.3, 195.5        & \nodata                                                            \\[3pt]
Time of periastron $T_0=t_{\rm ref}-P\frac{\lambda-\omega}{360\degree}$ (JD)& $2456380_{-60}^{+80}$            &      2456210, 2456520      & \nodata                                                            \\[3pt]
Mass ratio $q = M_{\rm comp}/M_{\rm host}$                                  & $0.0068_{-0.0016}^{+0.0015}$     &       0.0038, 0.0101       & \nodata                                                            
\enddata
\tablecomments{The $\chi^2$ of relative astrometry is 35.5 for separations and 32.3 for PAs, with 42 measurements for each. The $\chi^2$ of the \Hipparcos\ and \Gaia~DR2 proper motion differences is 1.09 for four measurements. For the parallax, we use a combination of the original and re-reduced \Hipparcos\ measurements re-weighted according to \citet{Brandt_2018}, $\varpi_{\rm HGCA} = 51.61\pm0.39$\,mas.}
\end{deluxetable*}

\begin{figure*}
\vskip -1.75 truein
\includegraphics[width=1.0\linewidth]{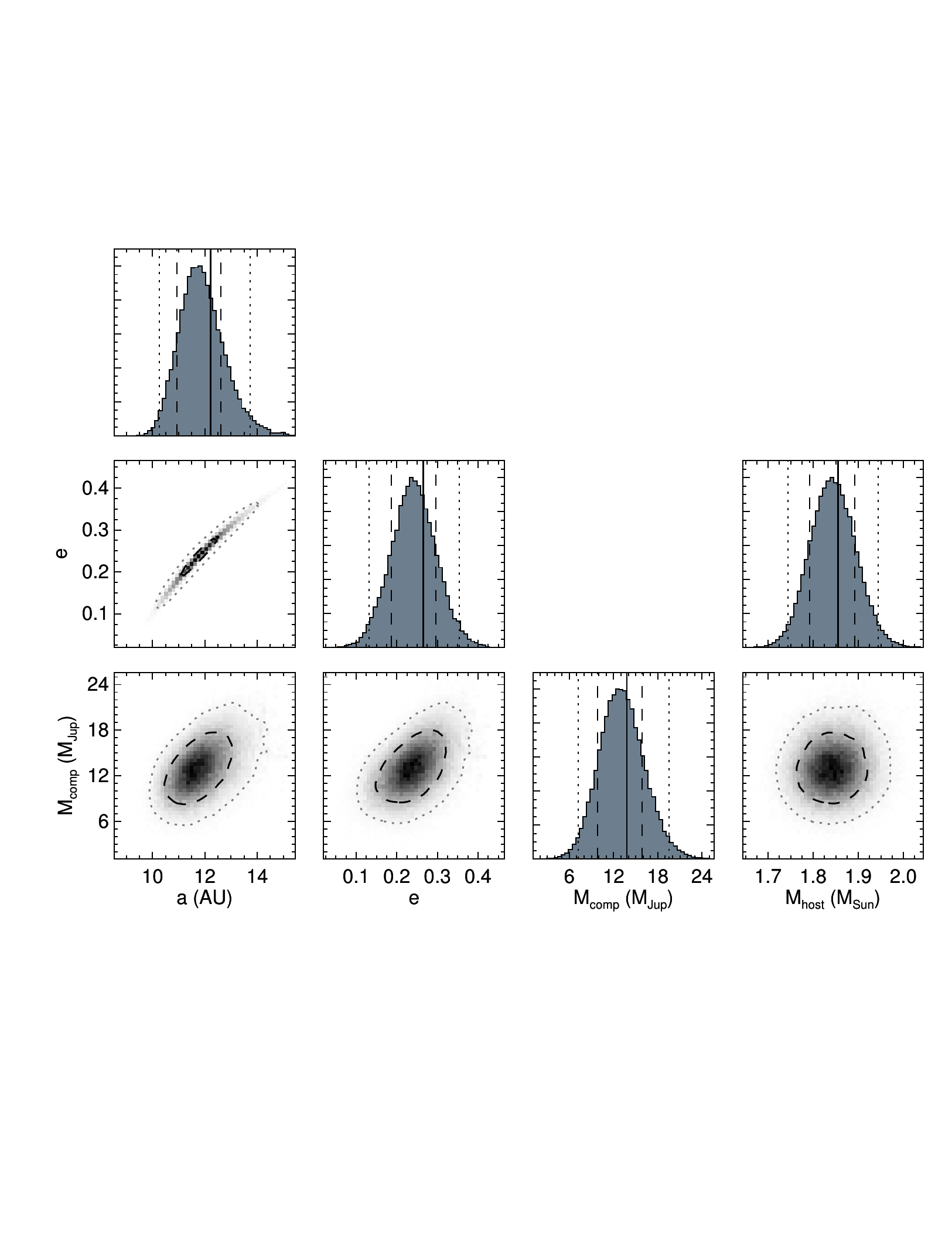}
\vskip -2.25 truein
\caption{Marginalized distributions for four orbital parameters (histograms) along with their joint posteriors (grayscale images with contours) for the \bpic\ system. In histograms, the thick solid lines indicate the highest likelihood orbit, and dashed and dotted lines show 1$\sigma$ and 2$\sigma$ ranges, respectively. In 2-d plots, the 1$\sigma$ and 2$\sigma$ areas of the joint posteriors are indicated by dark dashed contours and lighter dash-dotted contours, respectively. The strongest covariance is between eccentricity and semimajor axis such that both smaller, less eccentric and larger, more eccentric orbits are consistent with observations. Given the positive correlation between semimajor axis and period (not shown), less eccentric orbits also correspond to shorter orbital periods.}
\label{fig:posterior}
\end{figure*}

Posteriors of orbital parameters were determined using the parallel-tempering Markov chain Monte Carlo (PT-MCMC) ensemble sampler in \texttt{emcee~v2.1.0} \citep{2013PASP..125..306F} based on the algorithm described by \citet{2005PCCP....7.3910E}. We ran 30 temperatures and 100 walkers fitting for eleven parameters, including the masses of the host star ($M_{\rm host}$) and planet ($M_{\rm comp}$). Eight others define the orbit, including the zero point of the system velocity (RV$_{\rm zero}$) and the intrinsic RV jitter ($\sigma_{\rm jit}$). We also included parallax ($\varpi$) as a fitted parameter with a Gaussian prior based on the measured value. For the initial step, we drew random values according to our priors across all valid parameter space, where for log-flat priors we used bounds of 0.3--3.0\,\Msun\ in $M_{\rm host}$, 0.001--0.1\,\Msun\ in $M_{\rm comp}$, 1--100\,AU in $a$, and 0.3--300\,m\,s$^{-1}$ in $\sigma_{\rm jit}$. We used $3\times10^5$ steps in our PT-MCMC analysis, saving every 50th step of our chains. After ensuring that all walkers had stabilized in the mean and standard deviation of the posterior for each of the parameters we discarded all but the last $10^3$ samples as the burn-in portion yielding $10^5$ PT-MCMC samples across all walkers in the cold chain. Table~\ref{tbl:mcmc-BPIC} provides information on all our priors and posteriors, Figure~\ref{fig:orbit} shows our orbit fit compared to the input measurements, and Figure~\ref{fig:posterior} shows posteriors of astrophysically important parameters.

\section{Discussion} \label{sec:discussion}

Previous work has established key aspects of the orbit of \bpic~b, such as the viewing geometry of the nearly edge-on orbit and total system mass \citep[e.g.,][]{2012A&A...542A..41C,2014ApJ...794..158N,2014PNAS..11112661M,2014Natur.509...63S,2016AJ....152...97W}.
The reflex motion induced on \bpic\ by \bpic~b is this same orbit scaled down by the mass ratio. Because of stellar proper motion, detecting this reflex motion and obtaining a dynamical mass for \bpic~b requires measuring nonlinear perturbations on the motion of \bpic. In principle, RVs could determine a mass for \bpic~b, but the substantial RV jitter on such a young, active star ($\sigma_{\rm jit} = 269\pm6$\,m\,s$^{-1}$) hampers the measurement of the $\sim$100\,m\,s$^{-1}$ expected semiamplitude of the planet.

HGCA reports deviations from constant proper motion of only 1--2$\sigma$ for \bpic. Our joint fit of astrometry and RVs yields a mass posterior of $13\pm3$\,\Mjup\ (23\% uncertainty) for \bpic~b. This is not as precise as the value of $11\pm2$\,\Mjup\ (18\% uncertainty) from \citet{2018NatAs.tmp..114S} because we adopted cross-calibrated \Hipparcos\ and \Gaia~DR2 astrometry, which \citet{Brandt_2018} found requires error inflation of reported astrometric errors in both catalogs. Moreover, our analysis does not assume a host-star mass, although it broadly supports previous assumptions of 1.75\,\Msun\ with a remarkably precise model-independent mass of $1.84\pm0.05$\,\Msun. Our mass determination for \bpic~b is chiefly driven by the small offset between the \Hipparcos\ proper motion and the \Hipparcos-to-\Gaia\ positional difference, as was the case in \citet{2018NatAs.tmp..114S}, because the uncertainty in the \Gaia~DR2 proper motion is very large due to \bpic\ being saturated in \Gaia.

Combining our mass for \bpic~b with the luminosity of $\log(\Lbol/\Lsun) = -3.78\pm0.03$\,dex determined by \cite{2015ApJ...815..108M} we calculate upper and lower limits on the substellar cooling age from the hot-start evolutionary models of \citet{2008ApJ...689.1327S}. We use the same method described in \citet{2017ApJS..231...15D}, with a uniform prior in age, our orbit posterior as the prior on mass, and rejection-sampling on \Lbol\ to select models consistent with \bpic~b. The posterior on the age is wide, as expected given the low-precision mass. The 3$\sigma$ confidence interval on the cooling age ranges from 7--65\,Myr. Combining the 7\,Myr lower limit with external age information for \bpic\ and its eponymous young moving group directly determines the amount of time that could have elapsed between the formation of the host star and planet. Adopting the \bpic\ moving group age of $22\pm6$\,Myr from \citet{2017AJ....154...69S}, we thus find an upper limit of $15\pm6$\,Myr in the difference between the times of formation (a.k.a.\ $t = 0$) for \bpic~b and its host star. This is not particularly constraining on theory, but improved precision in the mass of \bpic~b in the future will result in stronger tests of the timescale of giant planet formation.

Our dynamical mass of $13\pm3$\,\Mjup\ for \bpic~b is broadly consistent with hot-start formation models, as these predict a mass of $13.0^{+0.4}_{-0.3}$\,\Mjup\ at $22\pm6$\,Myr \citep{2018AJ....156...57D}. The high-mass end of our posterior is also consistent with warm-start models. Our 2$\sigma$ upper limit on the mass of \bpic~b is 19.5\,\Mjup\ (0.019\,\Msun). Interpolating hot-start evolutionary tracks from \citet{2008ApJ...689.1327S}, an object of this mass should have a luminosity of $\log(\Lbol/\Lsun) = -3.1$\,dex at an age of 22\,Myr. The actual luminosity of \bpic~b is 0.7\,dex (1.7\,mag) fainter than this. This decrement corresponds to the intermediate range of 10-\Mjup\ warm-start models from \citet[][see their Figure~9]{2012ApJ...745..174S} and is highly inconsistent with cold-start models that are $\approx$5\,mag fainter than hot-start tracks at 22\,Myr. However, warm-start models would need to be computed beyond 10\,\Mjup\ for a more accurate appraisal.

Our relative orbit for \bpic~b is consistent with past work within the uncertainties, but our posteriors are notably lacking any near-circular orbits, with $e<0.1$ excluded at $>2\sigma$. Previous work was generally consistent with eccentricities up to 0.1--0.2 but preferred more circular orbits, unlike our orbit fit ($e=0.24\pm0.06$). Based on tests using various subsets of the relative astrometry, we find that our results are simply the consequence of combining all published measurements in a joint fit.  Recent results from \citet{2016AJ....152...97W} did not have access to the VLT/SPHERE measurements, and \citet{2018arXiv180908354L} used only VLT astrometry in their analysis. We note that our choice to exclude eight relative astrometry outliers out of 50 measurements does not significantly impact this result. We ran an identical PT-MCMC using all 50 measurements, with errors of 4\,mas and 0$\fdg$3 added in quadrature to all separations and PAs in order to make a reasonable $\chi^2$. All parameter posteriors were very similar, including a slightly higher eccentricity of $0.28\pm0.06$.

The strong preference of our fit for non-circular orbits has implications for the origin of \bpic~b and its history of dynamical interactions with the disk. While the focus of the literature has been on the planet--disk interaction as a means to explain the disk warp \citep[e.g.,][]{2011ApJ...743L..17D}, the eccentricity may help explain other observations \citep{2015ApJ...800..136A,2015ApJ...811...18M,2016AJ....152...97W}. 
The higher eccentricity for the planet ($e\sim0.2$) is also consistent with the exo-comet hypothesis put forth to explain the occasional absorption features in the host star's spectrum. \citet{2001A&A...376..621T} note that the frequency of observed events is well explained by the excitation of cometary bodies in a 3:1 resonance with a massive perturber at roughly 10\,AU. However, the location from which these comets would be launched lies somewhat inside the inner edge of the disk as fit by \citet{2015ApJ...811...18M}. Simulations of the planet--disk interaction at lower values of $e$ have not successfully explained the observed inner edge and disk scale height \citep{2015ApJ...811...18M,2015ApJ...815...61N}. These authors also note that a low eccentricity ($e<0.1$) is unable to account for the observed northeast--southwest brightness asymmetry in the disk. Although several authors have proposed a possible unseen second planet to explain these features \citep{2015ApJ...800..136A,2015ApJ...811...18M,2015ApJ...815...61N,2016AJ....152...97W}, further modeling with \bpic~b alone at a higher eccentricity may be warranted.

While the higher eccentricity for \bpic~b may help explain some present-day disk observations, it is consistent with a wide range of past formation scenarios. Giant planets are thought to form on relatively circular orbits due to efficient damping in the natal protoplanetary disk \citep{2011ARA&A..49..195A}, but many mechanisms can subsequently pump their eccentricities. High eccentricities can easily be generated by secular, resonant, or scattering interactions with a massive perturber in the form of another planet or nearby stars \citep{1997Natur.386..254H,1998ApJ...508L.171L,1996Sci...274..954R,2008ApJ...686..621F}. The perturber responsible for \bpic~b's eccentricity need not remain in the system and be observable today; it could have been ejected by a strong scattering event. As noted above, the presence of a second planet is favored in some models to explain disk structures. Detailed analysis of long-term RV monitoring excludes much of the parameter space for additional planets \citep{2018A&A...612A.108L}, and while absolute astrometry can potentially rule out more planets \citep{2018arXiv181108902K}, the fact that \bpic~b is only marginally detected in current observations complicates the interpretation of additional astrometric signals due to more planets.

A second massive planet is not, however, required to generate an eccentricity as high as $e=0.2$--0.3. In principle, migration of a very massive planet like \bpic~b through a massive gas disk \citep[e.g.,][]{2001A&A...366..263P,2018MNRAS.474.4460R} or even a planetesimal disk \citep{1998Sci...279...69M} can generate substantial eccentricity growth. While more modest-mass planets have their eccentricities damped by the disk, planets with masses $\gtrsim10\,\Mjup$ can have their eccentricities pumped through interaction with the 3:1 outer Lindblad resonance: the planet excites eccentricity in the disk, which back-reacts to excite eccentricity in the planet \citep{2012ARA&A..50..211K}. \citet{2001A&A...366..263P} explicitly predict that a massive eccentric planet inside a disk cavity is a natural outcome of this process. A key implication of this mechanism is that \bpic~b formed exterior to its current orbit.

\setlength{\tabcolsep}{0.047in}
\begin{deluxetable*}{lcccccccccccc}
\tablecaption{Predicted Future Astrometry and Radial Velocities for $\beta$ Pic b \label{tbl:future}}
\tablewidth{0pt}
\tablehead{
\colhead{Epoch}                      &
\colhead{}                           &
\multicolumn{3}{c}{Separation (mas)} & 
\colhead{}                           &
\multicolumn{3}{c}{PA (\degree)}     & 
\colhead{}                           &
\multicolumn{3}{c}{$\Delta{\rm RV}$ (\kms)} \\
\colhead{}                           &
\colhead{}                           &
\colhead{$e=0.10$}            &
\colhead{$e=0.20$}            &
\colhead{$e=0.30$}            &
\colhead{}                           &
\colhead{$e=0.10$}            &
\colhead{$e=0.20$}            &
\colhead{$e=0.30$}            &
\colhead{}                           &
\colhead{$e=0.10$}            &
\colhead{$e=0.20$}            &
\colhead{$e=0.30$}            }
\startdata
2019 Jan 1 &  &     $176\pm 3$ &     $173\pm 3$ &     $169\pm 3$ & & $28.21\pm0.31$ & $27.97\pm0.33$ & $27.79\pm0.35$ & & \phn$ 2.92\pm0.23$ & $ 1.24\pm0.20$ & $-0.25\pm0.19$\phs \\
2020 Jan 1 &  &     $301\pm 3$ &     $299\pm 3$ &     $296\pm 3$ & & $29.81\pm0.20$ & $29.63\pm0.21$ & $29.49\pm0.22$ & & \phn$ 5.76\pm0.29$ & $ 3.68\pm0.24$ & $ 1.87\pm0.22$     \\
2021 Jan 1 &  &     $407\pm 3$ &     $412\pm 3$ &     $413\pm 3$ & & $30.50\pm0.15$ & $30.32\pm0.17$ & $30.19\pm0.18$ & & \phn$ 8.07\pm0.32$ & $ 5.68\pm0.27$ & $ 3.61\pm0.25$     \\
2022 Jan 1 &  &     $490\pm 3$ &     $508\pm 3$ &     $517\pm 3$ & & $30.92\pm0.13$ & $30.73\pm0.14$ & $30.59\pm0.15$ & & \phn$ 9.78\pm0.33$ & $ 7.24\pm0.28$ & $ 5.00\pm0.26$     \\
2023 Jan 1 &  &     $545\pm 6$ &     $585\pm 4$ &     $608\pm 3$ & & $31.24\pm0.11$ & $31.02\pm0.12$ & $30.86\pm0.13$ & &     $10.89\pm0.30$ & $ 8.41\pm0.28$ & $ 6.10\pm0.26$     \\
2024 Jan 1 &  & \phn$571\pm10$ &     $642\pm 6$ &     $684\pm 4$ & & $31.51\pm0.11$ & $31.25\pm0.11$ & $31.06\pm0.12$ & &     $11.40\pm0.24$ & $ 9.22\pm0.26$ & $ 6.95\pm0.26$     \\
2025 Jan 1 &  & \phn$567\pm16$ &     $679\pm 9$ &     $747\pm 6$ & & $31.78\pm0.10$ & $31.44\pm0.10$ & $31.23\pm0.11$ & &     $11.29\pm0.17$ & $ 9.70\pm0.22$ & $ 7.58\pm0.25$     \\
2026 Jan 1 &  & \phn$533\pm23$ & \phn$695\pm13$ &     $794\pm 8$ & & $32.06\pm0.11$ & $31.62\pm0.10$ & $31.37\pm0.11$ & &     $10.59\pm0.18$ & $ 9.86\pm0.18$ & $ 8.02\pm0.23$     \\
2027 Jan 1 &  & \phn$471\pm30$ & \phn$690\pm18$ & \phn$827\pm11$ & & $32.40\pm0.14$ & $31.80\pm0.10$ & $31.51\pm0.10$ & & \phn$ 9.30\pm0.31$ & $ 9.73\pm0.14$ & $ 8.30\pm0.20$     \\
2028 Jan 1 &  & \phn$383\pm38$ & \phn$664\pm24$ & \phn$845\pm15$ & & $32.88\pm0.22$ & $31.99\pm0.10$ & $31.63\pm0.10$ & & \phn$ 7.44\pm0.52$ & $ 9.31\pm0.13$ & $ 8.42\pm0.17$     
\enddata
\tablecomments{Computed from subsets of our posterior selected by rejection sampling using Gaussian eccentricity priors with $\sigma_e = 0.01$.}
\end{deluxetable*}

Given the still limited observational coverage of the $\approx$30-year orbit of \bpic~b, and a particular lack of data on the northeastern side, its eccentricity is still relatively uncertain.  In Table~\ref{tbl:future}, we provide predicted astrometry and RVs for three representative eccentricities from our PT-MCMC posterior ($e=0.1$, 0.2, and 0.3). In the near term, the RV of \bpic~b is the most discriminating between different eccentricities, but after a few years separation measurements will cleanly define the orbit. Lower eccentricity orbits predict smaller separations in the next decade and a more imminent turnaround toward decreasing separation.

\acknowledgments

We thank the referee for a thoughtful and timely review.
This work has made use of data from the European Space Agency mission \Gaia\ (\url{https://www.cosmos.esa.int/gaia}), processed by the Gaia Data Processing and Analysis Consortium (DPAC, \url{https://www.cosmos.esa.int/web/gaia/dpac/consortium}). Funding for the DPAC has been provided by national institutions, in particular the institutions participating in the Gaia Multilateral Agreement.  
T.J.D.\ acknowledges research support from Gemini Observatory.
T.D.B.\ gratefully acknowledges support from the Heising-Simons foundation and from NASA under grant \#80NSSC18K0439.
Our research has employed NASA ADS; SIMBAD; VizieR; and J.~R.~A.\ Davenport's IDL implementation of the cubehelix color scheme \citep{2011BASI...39..289G}.


\end{document}